\documentclass[superscriptaddress, reprint,amsmath, amssymb, aps, pra, 10 pt]{revtex4-2}
\usepackage{multirow}
\usepackage{xcolor}
\usepackage{makecell}
\usepackage{graphicx}% Include figure files
\usepackage{dcolumn}% Align table columns on decimaluminum point
\usepackage{bm}% bold math

\begin{document}
\preprint{APS/TEOP neon on Al}

\title{Bremsstrahlung induced atomic processes}
%\title{Laboratory observations of unusually strong secondary ionization by the bremsstrahlung radiation}% Force lineonbreaks with \\
%\title{Observation of two-electron one-photon transitions and collisional-radiative process in collision of 1.8-2.1 MeV neon on aluminum}% Force lineonbreaks with \\
\author{Shashank Singh}
%\altaffiliation{Present address: IR Thermography and Charge Exchange Section, Plasma Diagnostic Division, Institute far Plasma Research, Ahmedabad-382428, India.}%Lines break automatically or can be forced with \\
\affiliation{Department of Physics, Panjab University, Chandigarh-160014, India.
}%
\author{Narendra Kumar}
\affiliation{School of physical sciences, JNU, New Delhi-110067, India}
\author{Soumya Chatterjee}
\affiliation{Department of Physics, Brainware University, Barasat, Kolkata-700125, India}
\author{Manpreet Kaur}
\affiliation{Department of Physics, University Institute of Sciences, Chandigarh University, Gharuan, Mohali, Punjab 140413, India}
%\author{Prashant Sharma}
%\affiliation{Prashant, please put your affiliation here}
\author{Deepak Swami}
\affiliation{Inter-University Accelerator Centre, Aruna Asaf Ali Marg, Near Vasant Kunj, New Delhi-110067, India.}%
\author{Alok Kumar Singh Jha}
\affiliation{School of physical sciences, JNU, New Delhi-110067, India}
\author{Mumtaz Oswal}%
\affiliation{%
 Department of Physics, DAV College, Sector 10, Chandigarh-160011, India.
}%
\author{K. P. Singh}
\affiliation{Department of Physics, Panjab University, Chandigarh-160014, India.
}%
%\affiliation{
\author{T. Nandi}
\thanks {Email:\hspace{0.0cm} nanditapan@gmail.com (corresponding author)}
\affiliation{Department of Physics, Ramakrishna Mission Vivekananda Educational and Research Institute, PO Belur Math, Dist Howrah 711202, West Bengal, India}

%\altaffiliation{%superannuated from
%Inter-University Accelerator Centre, Aruna Asaf Ali Marg, Near Vasant Kunj, New Delhi-110067, India.}
%}\email{nanditapan@gmail.com}


\date{\today}% It is always \today, today,

\begin{abstract}
The observed spectra in the collisions of neon (Ne) projectiles of 1.8 and 2.1 MeV with an aluminum target (Al) have been successfully segregated from strong bremsstrahlung backgrounds and then analyzed by comparing the transition energies and rates with the theoretical predictions of the flexible atomic structure code and the general-purpose relativistic atomic structure package. The spectra contain $K_\alpha$, $K^h_\alpha$, and $K_{\alpha\alpha}$ lines.  The $K_{\alpha\alpha}$ emissions are due to two-electron one-photon transitions. Interestingly, the $K_{\alpha\alpha}$ lines in projectile ions are only seen with 1.8 MeV energy. In contrast, the $K_{\alpha\alpha}$ lines in the target ions are also well observed with 2.1 MeV energy. Surprisingly, the Al $K$ x-ray line intensities are strongly suppressed, and the $K_{\alpha\alpha}$ line intensities are unexpectedly enhanced. The underlying physical process is found to be the photoionization caused by intense bremsstrahlung radiation. This photoionization process converts most of the singly ionized $K$ shell states ($\approx$82\% at 2.1 MeV) to doubly ionized $K$-shell states. This phenomenon is silently present on many occasions. We take some of such events to validate this remarkable finding. This bremsstrahlung radiation induced secondary ionization process stands as an eye opening incidence to the plasma physics, astronomy and astrophysics communities; may revolutionize these fields of research.
\end{abstract}
\maketitle

%\tableofcontents
\section{Introduction}
\begin{figure} 
\includegraphics[width=85mm,height=48mm,scale=01.0]{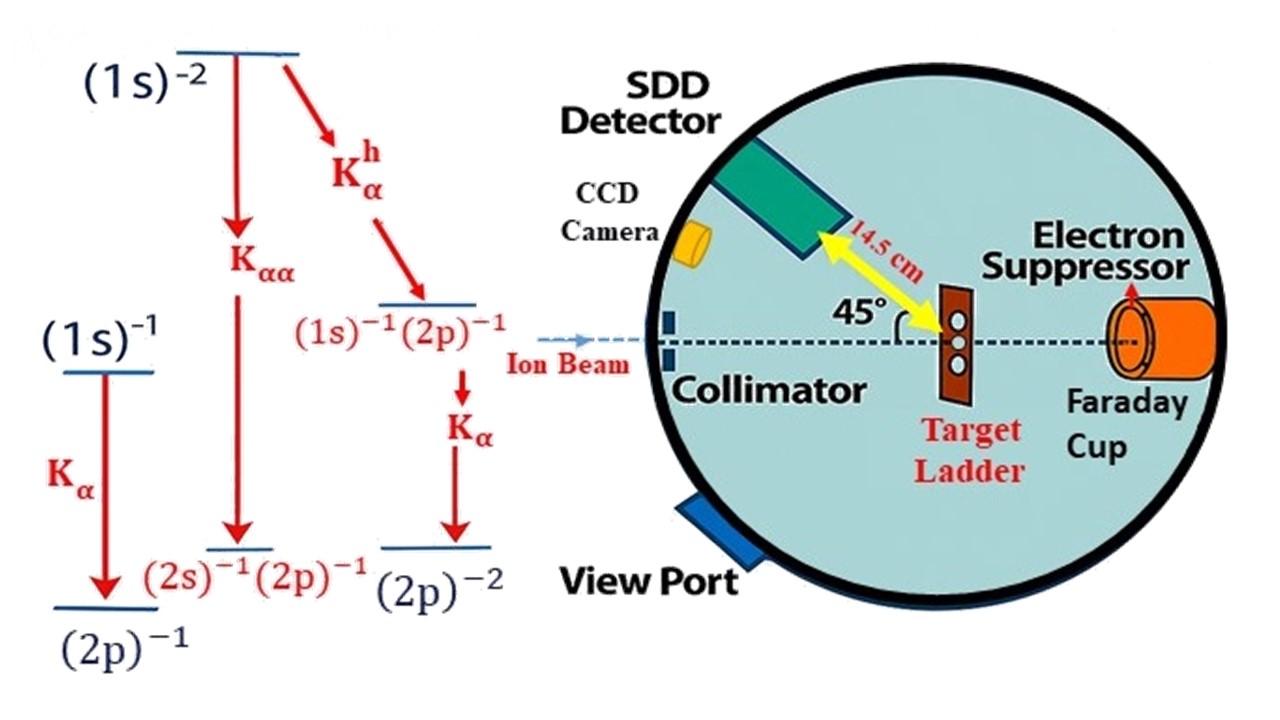}
\caption{\label{fig:1} Schematic of the two-electron one-photon transition (left) and experimental setup for the experiment (right).}
\end{figure}

In heavy-ion-atom collision processes, the creation of multiple vacancies in different inner shells simultaneously is a common phenomenon. Sometimes, the $K$ shell of the target atom or projectile ions may be fully ionized. Normally, two vacancies are filled in two steps, emission of hypersatellites and satellite x-ray lines are emitted one after another, as shown pictorially in Fig. \ref{fig:1}. Nevertheless, according to the prediction of Refs. \cite{Heisenberg1925quantentheorie,condon1930theory,goudsmit1931many}, it is even possible that occasionally both vacancies of the $K$ shell may be filled by a correlated jump of two electrons and only one photon is emitted. The energy of the two-electron one-photon (TEOP) transition ($K^{-2}-L^{-2}$) can be estimated as the sum of the transition energies of the hypersatellite ($K^{-2}-K^{-1}L^{-1}$) and satellite transition ($K^{-1}L^{-1}-L^{-2}$) energies. The schematic diagram of the TEOP transition is shown in Fig. \ref{fig:1}.
\par
%%%%%%%%%%%%%
%%%%%%%%%%%%%%%
\citet{wolfli1975two} reported the TEOP transitions, for the first time, for the Ni-Ni and Ni-Fe systems at beam energies $\approx$ 0.8 MeV/u. \citet{aaberg1976origin} showed that the TEOP transitions are due to the $1s^{-2}\rightarrow2s^{-1}2p^{-1}$ electric dipole transition by the calculations of the corresponding Hartree-Fock energy and transition probability. Various authors \citet{wolfli1976calculation}, \citet{nagel1976two}, \citet{aaberg1976origin} and \citet{knudson1976energies} reported the theoretical prediction of the TEOP transition energies in different systems to compare with the observed energies. In these works, Hartree-Fock energies have been consistent with the observed transition energies. \citet{stoller1977two} measured the TEOP transitions for Al-Al, O-Ca, Ca-Ca, Fe-Fe, Fe-Ni, Ni-Fe, and Ni-Ni systems at beam energies between 24 and 40 MeV. In this study, the cross sections of characteristic transitions and TEOP transitions were determined. In addition, branching ratios between the one-electron one-photon (OEOP) and TEOP transitions in doubly ionized $K$ shells were determined to compare them with various theoretical predictions and were found to be qualitatively consistent. In addition to ion-atom collisions, \citet{auerhammer1988two} studied the TEOP transition in the Al-target with electrons of energy 20 keV using a high resolution crystal spectrometer. 
%%%%%%%%%%%%%%%%%%%
\par
\citet{mukherjee1990branching} evaluated the branching ratio between the OEOP and TEOP transitions in various atomic systems that have a doubly ionized $K$ shell.  Subsequently, \citet{mukherjee1997two} reported the theoretical estimation of excitation energies and transition probabilities for TEOP transitions for the inner shell ionized atoms for He-like Ne, Na, Mg, Al, Si, P, S, Cl, and Ar ions. Both the OEOP and the TEOP transitions in the atomic systems of fully vacant $K$-shell are very sensitive to the Breit interaction, quantum electrodynamics (QED) and electron correlations \cite{diamant2000cu,briand1976two,kadrekar2010two}. Experimental attempts have also been made to study such physics aspects by monochromatic synchrotron radiation on solid targets. Recently, \citet{togawa2020observation} observed the TEOP spectra by incident photons on ions and compared the transition energies well with the flexible atomic code (FAC) calculations \cite{gu2008flexible}.
%%%%%%%%%%%%%%%%%%%%
\par
The study of TEOP with low energy heavy ions is very important for highly stripped ions of He- and Li-like ion sequences \cite{andriamonje1991two,mukherjee1995interpretation} as such ions are present in solar corona, flares \cite{kato1997highly} and laboratory Tokamak plasmas \cite{beiersdorfer2015highly}, but laboratory investigations on such systems are very sparse. We intended to conduct a thorough experimental investigation of TEOP radiations using low-energy Ne ions on an Al-target so that both the highly charged projectile and target ions could be used for the spectral analysis. We analyzed the spectra successfully, but encountered an unusual truth that two-electron one-photon line intensities are stronger than one-electron one-photon line intensities. A thorough search for the physical origin behind such an unexpected occurrence provides us with a new phenomenon: the bremsstrahlung radiation-induced photoionization process converts most of the singly-ionized $K$ shell states to doubly ionized $K$-shell states. Critical analysis and best interpretation of intriguing results will be portrayed in great detail.
\par
%%%%%%%%%%
\section{experimental details}
This experiment was carried out at the Low Energy Ion Beam Facility, IUAC, New Delhi. In this experiment, we used the Ne$^{6+}$ beam of energies between 1.8 and 2.1 MeV. The ion beam was extracted from an electron cyclotron resonance ion source placed on a high voltage deck of 400 kV. The vacuum chamber was located on the $75^\circ$ beam line. The schematic of the scattering chamber with the experimental setup is shown in Fig.\ref{fig:1}. A high vacuum around $2\times$10$^{-6}$Torr was maintained in the chamber. To detect x-rays, a silicon drift detector (SDD) was mounted at $135^\circ$ with respect to the beam direction. The SDD detector specification (KETEK AXAS-A) is as follows: active area = 20 $mm^2$, thickness of the beryllium window = $8\mu m$ and full width at half maxima = 124.2 eV for Mn $K_\alpha$ x-rays. The target ladder was mounted $90^\circ$ with respect to the beam direction, which holds five targets at a time, and a target was brought along the beam axis by a target manipulator. Spectroscopically pure Al-targets were $\approx$$~20~\mu g/cm^2$  measured by the energy loss of the $\alpha$ beam from an $^{241}$ Am source. The distance between the target ladder and the detector collimator (2.87 mm dia) was 14.5 cm. To measure the incident charge on the target, one Faraday cup was mounted behind the target and attached to a current integrator. A 5-mm-diameter beam collimator was placed at the entrance of the chamber.
\par
\begin{figure*}
\includegraphics[width=170mm,height=110mm]{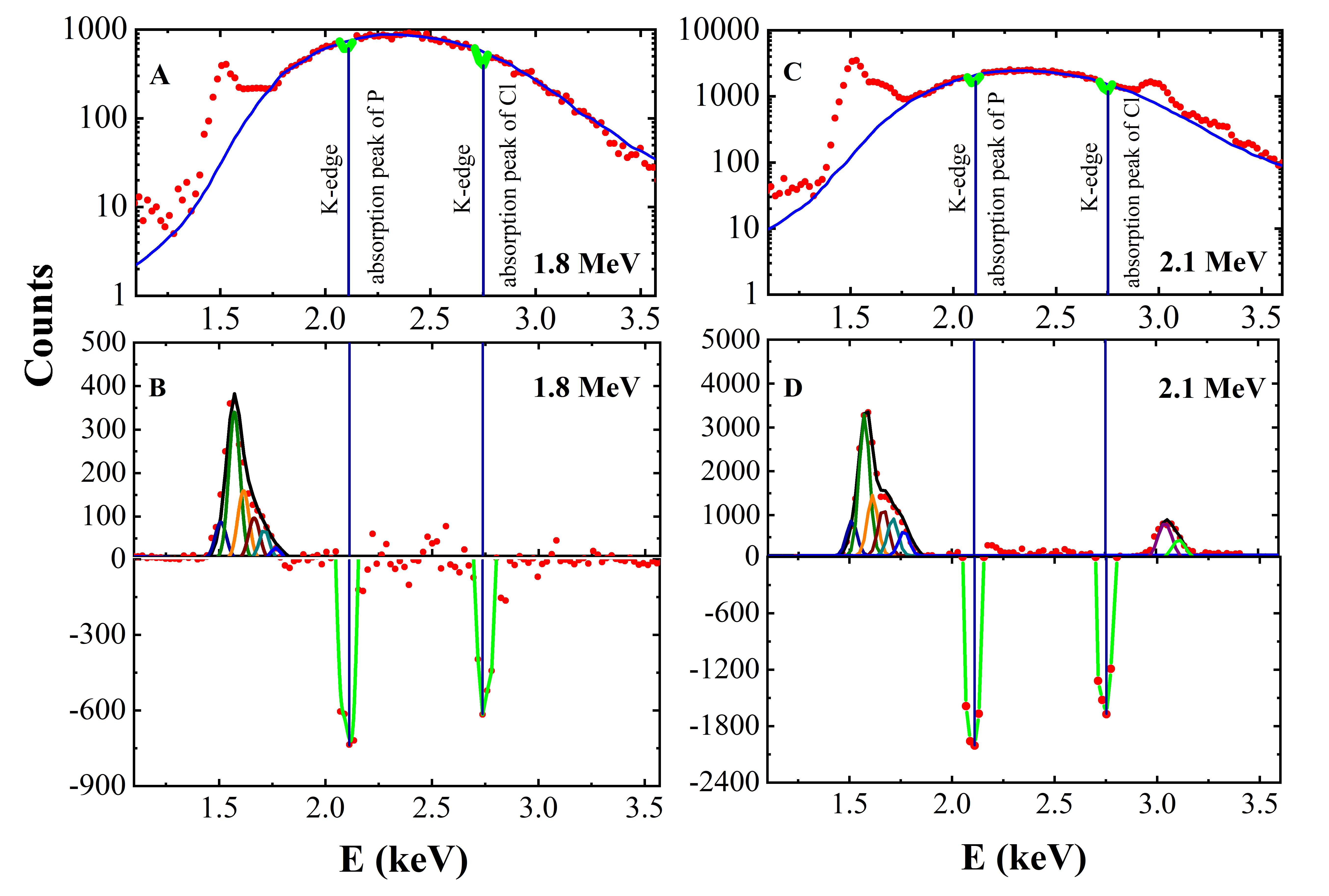}
          \caption{The lines in blue (A and C) represent the background spectra of the Ne ions on Al-target at 1.8 (A) and 2.1 MeV (C), respectively. The background subtracted spectra (B and D) are fitted with Gaussian peaks. The x-axis represents the x-ray energy in keV and y-axis the counts. While the vertical lines in A as well as C are marked as the $K$-edge absorption lines of phosphorous (P) and chlorine (Cl), which are used to refine the spectrum calibrations.}
\label{fig:expt}
\end{figure*}
%%%%%%%%%%%%%%%%%%%%%%%%%%%%%%
\begin{figure}[htbp]
    \centering
    \includegraphics[width=90mm,height=90mm]{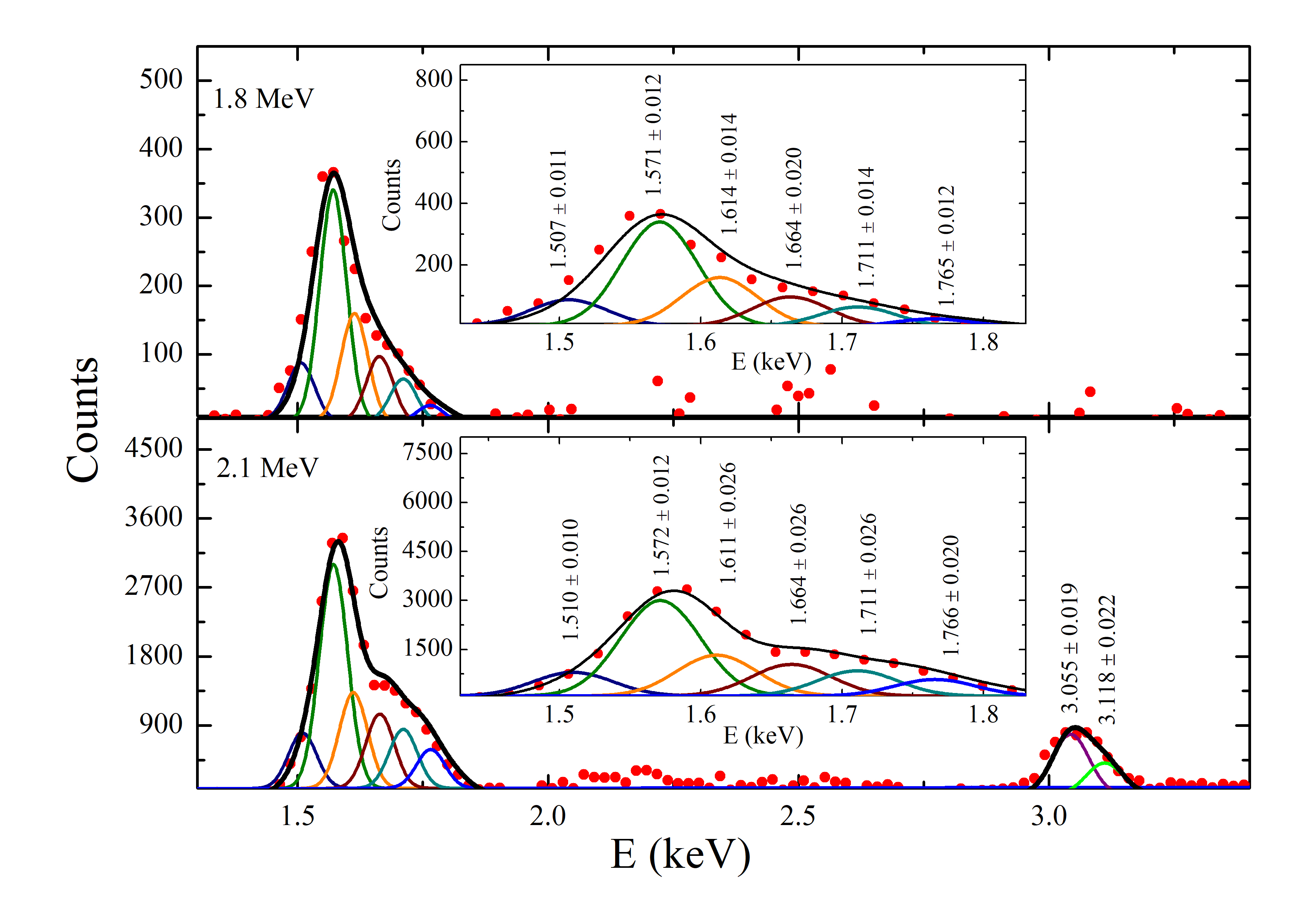}
    \caption{The signal regions shown in Fig.\ref{fig:expt} are more clearly displayed first and zoomed in further in the insets to label the fitted spectral lines.}
    \label{AA}
\end{figure}
%\vspace{-100mm}
\par
The thickness of the mylar window for the SDD detector in the vacuum chamber used was $10\mu$m. X-ray energy calibration was done before and after the experiment using a $^{241}$ Am radioactive source. One position of the target on the target ladder was kept empty to study whether the beam hits the target frame or not. Furthermore, it also allowed us to compare the number of incident particles without the target and with the target in position using the Faraday cup. At another position on the target ladder, a quartz crystal was placed to visualize the beam position. The online data was acquired using a CAMAC-based software called FREEDOM \cite{freedomacquisition}.

%%%%%%%%%%%%%%%
\section{Theoretical methods}
\par
The present calculations are performed using the fully relativistic Dirac-Fock multiconfiguration method (MCDF) \cite{grant1980atomic} implemented in the general purpose relativistic atomic structure package (GRASP) and the relativistic configuration interaction (RCI) technique imposed in flexible atomic code (FAC) \cite{gu2008flexible}. An elaborate description of these two theories has been presented in \citet{kumar2023relativistic} and \citet{10.1093/mnras/stad2053}, respectively. We considered all ionic states in both the cases of Ne and Al. To the best of our knowledge, such an effort has hardly been seen in the literature. We considered a number of configurations in each ionic state to ensure the convergence of the results. This means that both the transition energies and the rates do not vary much either with additional configurations or with the length and velocity gauges. We must note that though the output of GRASP and FAC code comes in LS and JJ couplings, respectively, they appear in the sequence. Hence, comparing the outputs is simple and we may use either LS or JJ coupling schemes. In the present case, we choose the former scheme for further use.  Another point is that in case multiple theoretical candidates exist within the experimental uncertainty range, the closest energy not only matches the necessary criterion but also the highest transition rate is considered. Furthermore, the observed lines are assigned in such a manner that the difference between two consecutive lines measured is consistently in agreement with that of the calculated ones.
\par
 %%%%%%%%%%%%%%%%%%%%%%%%%%%%%%%%%%%%%%%%%%%%%%%%%%%%%%%%%
\begin{table*}
\caption{Testing the present calculation using the FAC and GRASP codes by comparing with the earlier theoretical as well as experimental results of transition energies (E) and rates (A). Two different transition energies shown with the same ref. \cite{kadrekar2010two} represent values with and without the higher order correlations.}
\label{tab:theo}
\resizebox{18 cm}{!}{
\begin{tabular}{|c|c|c|c|c|c|cc|c|c|c|}
\hline
\multirow{2}{*}{El.} & \multirow{2}{*}{Trans.} & \multirow{2}{*}{Initial State}                  & \multirow{2}{*}{Final State}                 & \multirow{2}{*}{E(FAC)} & \multirow{2}{*}{E(GRASP)} & \multicolumn{2}{c|}{Previous Results (E)}                                                                                       & \multirow{2}{*}{A(FAC)} & \multirow{2}{*}{A(GRASP)} & \multirow{2}{*}{Prev. Res. (A)} \\ \cline{7-8}
                     &                         &                                                 &                                              &                      &                        & \multicolumn{1}{c|}{Theory}                                                                                   & Experiment &                          &                            &                               \\ \hline
Ne                  & $K_{\alpha\alpha}$      & $1s^02s^22p^6 \quad ^1S_0^e$                    & $1s^22s2p^5 \quad ^1P_1^o$                   & 1.765                & 1.758                  & \multicolumn{1}{c|}{1.808 \cite{mukherjee1997two}, 1.762~\cite{gavrila1976calculation}}                                      &            & 5.930E9                 & 7.045E9                   & 9.237E9~\cite{mukherjee1997two}                       \\ \hline
Ne                  & $K_{\alpha\alpha}$      & $1s^02s^22p^5 \quad ^2P_{3/2}^{o}$              & $1s^22s2p^4     \quad ^2P_{3/2}^e$           & 1.789                & 1.793                  & \multicolumn{1}{c|}{}                                                                                         &            & 4.014E9                 & 3.509E9                   &                               \\ \hline
Ne                  & $K_{\alpha\alpha}$      & $1s^02s^22p^4 \quad ^1D_2^e$                    & $1s^22s2p^3 \quad ^1P_1^o$                   & 1.797                & 1.804                  & \multicolumn{1}{c|}{}                                                                                         &    1.8~\cite{andriamonje1991two}        & 8.098E9                 & 9.280E9                   &                               \\ \hline
Ne                  & $K_{\alpha\alpha}$      & $1s^02s^22p^3 \quad ^2P_{3/2}^{o}$              & $1s^22s^22p     \quad ^2P_{3/2}^{e}$         & 1.829                & 1.832                  & \multicolumn{1}{c|}{}                                                                                         &            & 6.642E9                 & 7.424E9                   &                               \\ \hline
Ne                  & $K_{\alpha\alpha}$      & $1s^02s2p \quad ^1P_{1}^{o}$                    & $1s^2 \quad ^1S_{0}^{e}$                     & 1.927                & 1.926                  & \multicolumn{1}{c|}{1.924 \cite{mukherjee1997two}}                                                                                         &            & 1.066E9                 & 3.004E9                   & 4.514E9~\cite{mukherjee1997two}                       \\ \hline
Al                   & $K_{\alpha}$            & $1s2s^22p^6 3s^23p \quad  ^3P_0^o$            & $1s^22s^22p^5 3s^23p  \quad ^3D_1^e$         & 1.487                & 1.492                  & \multicolumn{1}{c|}{1.486~\cite{winick2009x}}                                                &            & 1.258E13                & 1.546E13                  &                               \\ \hline
Al                   & $K_{\alpha}$            & $1s2s^22p^6 3s^2\quad        ^2S^e_{1/2}$       & $1s^22s^22p^53s^2\quad    ^2P^o_{3/2}$       & 1487                 & 1.495                  & \multicolumn{1}{c|}{}                                                                                         &            & 1.340E13                & 1.513E13                  &                               \\ \hline
Al                   & $K_{\alpha}$            & $1s2s^22p^6 \quad             ^2S^e_{1/2}$      & $1s^22s^22p^5\quad     ^2P^o_{3/2}$          & 1.489                & 1.493                  & \multicolumn{1}{c|}{}                                                                                         &            & 1.479E13                & 1.557E13                  &                               \\ \hline
Al                   & $K_{\alpha}$            & $1s2s^22p^5 \quad ^1P^o_1$                      & $1s^22s^22p^4 \quad  ^3P^e_2$                & 1.502                & 1.503                  & \multicolumn{1}{c|}{}                                                                                         & 1.497$\pm$ 0.001 \cite{auerhammer1988two}            & 2.738E13                & 2.833E13                  &                               \\ \hline
Al                   & $K_{\alpha}$            & $1s2s^22p^4 \quad ^4P^e_{5/2}$                  & $1s^22s^22p^3 \quad ^2D^o_{5/2}$             & 1.514                & 1.513                  & \multicolumn{1}{c|}{}                                                                                         & \begin{tabular}[c]{@{}c@{}}1.513$\pm$ 0.003 \cite{stoller1977two}\end{tabular}          & 1.463E13                & 1.284E13                  &                               \\ \hline 
Al                   & $K_{\alpha}$            & $1s2s^22p^3 \quad ^1D^o_{2}$                  & $1s^22s^22p^2 \quad ^1D^e_{2}$             & 1.529                & 1.530                  & \multicolumn{1}{c|}{}                                                                                         & \begin{tabular}[c]{@{}c@{}}\end{tabular}          & 3.081E13                & 3.144E13                  &                               \\ \hline
Al                   & $K_{\alpha}$            & $1s2s^22p^2 \quad ^2P^e_{3/2}$                  & $1s^22s^22p\quad ^2P^o_{3/2}$             & 1.548                & 1.548                  & \multicolumn{1}{c|}{}                                                                                         & \begin{tabular}[c]{@{}c@{}}\end{tabular}          & 2.991E13                & 3.058E13                  &                               \\ \hline
Al                   & $K_{\alpha}$            & $1s2s^22p \quad ^1P^o_{1}$                  & $1s^22s^2 \quad ^1S^e_{0}$             & 1.568                & 1.568                  & \multicolumn{1}{c|}{}                                                                                         & \begin{tabular}[c]{@{}c@{}}\end{tabular}          & 2.368E13                & 2.391E13                  &                               \\ \hline
Al                   & $K_{\alpha}$            & $1s2s2p \quad ^2P^o_{3/2}$                  & $1s^22s \quad ^2S^e_{1/2}$             & 1.580                & 1.580                  & \multicolumn{1}{c|}{}                                                                                         & \begin{tabular}[c]{@{}c@{}}\end{tabular}          & 2.453E13                & 2.516E13                  &                               \\ \hline
Al                   & $K_{\alpha}$            & $1s2p \quad ^1P^o_{1}$                  & $1s^2 \quad ^1S^e_{0}$             & 1.597                & 1.597                  & \multicolumn{1}{c|}{}                                                                                         & \begin{tabular}[c]{@{}c@{}}\end{tabular}          & 2.802E13                & 2.962E13                  &                               \\ \hline
Al                   & $K_{\alpha}$            & $2p \quad ^2P^o_{3/2}$                  & $1s \quad ^2S^e_{1/2}$             & 1.729                & 1.729                  & \multicolumn{1}{c|}{}                                                                                         & \begin{tabular}[c]{@{}c@{}}\end{tabular}          & 1.792E13                & 1.786E13                  &                               \\ \hline
Al                   & $K^h_{\alpha}$          & $1s^02s^22p^6 3s^23p    \quad ^2P_{1/2}^{o}$    & $1s^12s^22p^5 3s^2 3p\quad ^2D_{3/2}^{e}$    & 1.613                & 1.619                  & \multicolumn{1}{c|}{1.627 \cite{mukherjee1997two},  1.622~\cite{saha2009effect}}                                             & 1.611 $\pm$ 0.001 \cite{auerhammer1988two}           & 2.995E13                & 3.552E13                  & 1.547E13~\cite{mukherjee1990branching}                      \\ \hline
Al                   & $K^h_{\alpha}$          & $1s^02s^22p^6 3s^2    \quad ^1S^e_0$            & $1s^12s^22p^5 3s^2 \quad ^1P^o_1$            & 1.614                & 1.621                  & \multicolumn{1}{c|}{}                                                                                         &            & 5.187E13                & 6.188E13                  &                               \\ \hline
Al                   & $K^h_{\alpha}$          & $1s^02s^22p^6 3s    \quad ^2S^e_{1/2}$          & $1s^12s^22p^5 3s \quad ^2P^o_{3/2}$          & 1.614                & 1.626                  & \multicolumn{1}{c|}{}                                                                                         &            & 3.771E13                & 4.216E13                  &                               \\ \hline
Al                   & $K^h_{\alpha}$          & $1s^02s^22p^6 \quad ^1S^e_0$                    & $1s^12s^22p^5 \quad ^1P^o_1$                 & 1.615                & 1.627                  & \multicolumn{1}{c|}{}                                                                                         &            & 6.144E13                & 6.548E13                  &       1.275E13~\cite{mukherjee1997two}                        \\ \hline
Al                   & $K^h_{\alpha}$          & $1s^02s^22p^5 \quad ^2P^o_{3/2}$                & $1s^12s^22p^4 \quad ^4P^e_{3/2}$             & 1.627                & 1.634                  & \multicolumn{1}{c|}{}                                                                                         &            & 2.446E13                & 2.566E13                  &                               \\ \hline
Al                   & $K^h_{\alpha}$          & $1s^02s^22p^4 \quad ^1S^e_0$                    & $1s^12s^22p^3 \quad ^3S^o_1$                 & 1.645                & 1.649                  & \multicolumn{1}{c|}{}                                                                                         &    1.648 $\pm$ 0.001 \cite{stoller1977two}      & 5.028E13                & 5.163E13                  &                               \\ \hline
Al                   & $K^h_{\alpha}$          & $1s^02s^22p^3 \quad ^4S^o_{3/2}$                & $1s^12s^22p^2 \quad ^4P^e_{3/2}$             & 1.661                & 1.662                  & \multicolumn{1}{c|}{}                                                                                         &   1.668 $\pm$ 0.003 \cite{stoller1977two}        & 2.054E13                & 2.037E13                  &                               \\ \hline
Al                   & $K^h_{\alpha}$          & $1s^02s^22p^2 \quad ^3P^e_0$                    & $1s^12s^22p \quad ^3P^o_1$                   & 1.680                & 1.683                  & \multicolumn{1}{c|}{}                                                                                         &   1.684 $\pm$ 0.005 \cite{stoller1977two}       & 3.082E13                & 3.019E13                  &                               \\ \hline
Al                   & $K^h_{\alpha}$          & $1s^02s^22p^1 \quad ^2P^o_{1/2}$                & $1s^12s^2 \quad ^2S^e_{1/2}$                 & 1.695                & 1.694                  & \multicolumn{1}{c|}{}                                                                                         &            & 1.582E13                & 1.723E13                  &                               \\ \hline
Al                   & $K_{\alpha\alpha}$      & $1s^02s^22p^6 3s^23p \quad     ^2P^{o}_{1/2}$ & $1s^22s2p^5 3s^2 3p \quad     ^2D^{e}_{3/2}$ & 3.058                & 3.072                  & \multicolumn{1}{c|}{\begin{tabular}[c]{@{}c@{}}3.146 \cite{mukherjee1990branching}, 3.055 \cite{kadrekar2010two}\\  3.057 \cite{saha2009effect}, 3.056 \cite{martins2004relativistic}\end{tabular}} &       3.077$\pm$ 0.021~\cite{auerhammer1988two}     & 5.139E10                & 7.620E10                  & \multicolumn{1}{c|}{\begin{tabular}[c]{@{}c@{}} 4.043E10 \cite{mukherjee1990branching}\\ 3.248E10 \cite{kadrekar2010two}\\6.560E10 \cite{martins2004relativistic} \end{tabular}}                      \\ \hline
Al                   & $K_{\alpha\alpha}$      & $1s^02s^22p^6 3s^2 \quad     ^1S^{e}_{0}$       & $1s^22s2p^5 3s^2 \quad     ^3S^{o}_{0}$      & 3.059                & 3.057                  & \multicolumn{1}{c|}{3.060 \cite{kadrekar2010two} , 3.058 \cite{kadrekar2010two}}                                                                                         &            & 1.066E11                & 1.707E11                  &                               \\ \hline
Al                   & $K_{\alpha\alpha}$      & $1s^02s^22p^6 3s \quad     ^2S^{e}_{1/2}$       & $1s^22s2p^5 3s \quad     ^2P^{o}_{1/2}$      & 3.060                & 3.089                  & \multicolumn{1}{c|}{}                                                                                         &            & 1.478E10                & 1.353E10                  &                               \\ \hline
Al                   & $K_{\alpha\alpha}$      & $1s^02s^22p^6  \quad        ^1S^e_o$            & $1s^22s2p^5 \quad ^1P_1^o$                   & 3.064                & 3.058                  & \multicolumn{1}{c|}{}                                                                                         &            & 1.730E10                & 1.736E10                  &                               \\ \hline
Al                   & $K_{\alpha\alpha}$      & $1s^02s^22p^5 \quad  ^2P^o_{3/2}$               & $1s^2 2s2p^4 \quad  ^2S^e_{1/2}$             & 3.094                & 3.103                  & \multicolumn{1}{c|}{}                                                                                         &            & 9.259E9                 & 8.218E9                   &          1.628E10~\cite{mukherjee1997two}                     \\ \hline
Al                   & $K_{\alpha\alpha}$      & $1s^02s^22p^4      \quad      ^1D^e_2$          & $1s^22s2p^3 \quad      ^1P^1_0$              & 3.121                & 3.127                  & \multicolumn{1}{c|}{}                                                                                         &            & 1.758E10                & 2.022E10                  &                               \\ \hline
Al                   & $K_{\alpha\alpha}$      & $1s^02s^22p^3 \quad    ^2S^{o}_{3/2}$           & $1s^22s2p^2 \quad    ^2P^{e}_{3/2}$          & 3.164                & 3.168                  & \multicolumn{1}{c|}{}                                                                                         & 3.180 $\pm$ 0.020 \cite{stoller1977two}           & 1.873E10                & 2.865E10                   &                               \\ \hline
Al                   & $K_{\alpha\alpha}$      & $1s^02s^22p^2 \quad ^1S^e_{0}$                  & $1s^22s2p \quad ^1P^o_{1}$                   & 3.212                & 3.213                  & \multicolumn{1}{c|}{}                                                                                         &            & 9.466E10                & 9.796E10                  &                               \\ \hline
Al                   & $K_{\alpha\alpha}$      & $1s^02s^22p \quad ^2P^o_{1/2}$                  & $1s^22s \quad ^2S^e_{1/2}$                   & 3.257                & 3.257                  & \multicolumn{1}{c|}{}                                                                                         &            & 3.978E10                & 4.168E10                  &                               \\ \hline
Al                   & $K_{\alpha\alpha}$      & $1s^02s2p \quad ^1P_{1}^{0}$                    & $1s^2 \quad ^1S_{0}^{e}$                     & 3.304                & 3.302                  & \multicolumn{1}{c|}{3.296 \cite{mukherjee1997two}}                                                                                         &            & 2.257E9                 & 5.077E9                   & 7.822E9 \cite{mukherjee1997two}                       \\ \hline
\end{tabular}}
\end{table*}

\begin{table*}
\caption{\label{tab:Expt-1} Measured line energies with the 1.8 MeV Ne ions are listed and identified with the help of the theoretical calculations presented here as shown in Table \ref{tab:theo}. }
\resizebox{18cm}{!}{
\begin{tabular}{|l|l|l|l|l|l|l|l|l|}
\hline
El & Trans. & 1.8 MeV & Initial State & Final State & FAC & GRASP & A(FAC) & A(GRASP) \\ \hline

\multirow{4}{*}{Al} 
& $K^1_{\alpha}$ & 1.507$\pm$0.011 
& $1s2s^22p^5 \quad ^1P^o_1$ & $1s^22s^22p^4 \quad ^3P^e_2$ 
& 1502 & 1.503 & 2.738E13 & 2.833E13 \\ 
& & & $1s2s^22p^4 \quad ^4P^e_{5/2}$ & $1s^22s^22p^3 \quad ^2D^o_{5/2}$ 
& 1.514 & 1.513 & 1.463E13 & 1.284E13 \\ \cline{2-9}

& $K^2_{\alpha}$ & 1.571$\pm$0.012
& $1s2s^22p \quad ^1P^o_{1}$ & $1s^22s^2 \quad ^1S^e_{0}$ 
& 1.568 & 1.568 & 2.368E13 & 2.391E13 \\ 
& & & $1s2s2p \quad ^2P_{3/2}^{o}$ & $1s^22s \quad ^2S_{1/2}^{e}$ 
& 1.580 & 1.580 & 2.453E13 & 2.516E13 \\ \cline{2-9}

& $K^{h1}_{\alpha}$ & 1.614$\pm$0.014
& $1s^02s^22p^6 3s^2 3p \quad ^2P_{1/2}^{o}$ & $1s^12s^22p^5 3s^2 3p \quad ^2D_{3/2}^{e}$ 
& 1.613 & 1.619 & 2.995E13 & 3.552E13 \\ 
& & & $1s^02s^22p^6 3s^2 \quad ^1S^e_0$ & $1s^12s^22p^5 3s^2 \quad ^1P^o_1$ 
& 1.614 & 1.621 & 5.187E13 & 6.188E13 \\ 
& & & $1s^02s^22p^6 3s \quad ^2S^e_{1/2}$ & $1s^12s^22p^5 3s \quad ^2P^o_{3/2}$ 
& 1.614 & 1.626 & 3.771E13 & 4.216E13 \\ 
& & & $1s^02s^22p^6 \quad ^1S^e_0$ & $1s^12s^22p^5 \quad ^1P^o_1$ 
& 1.615 & 1.627 & 6.144E13 & 6.548E13 \\ 
& & & $1s^02s^22p^5 \quad ^2P^o_{3/2}$ & $1s^12s^22p^4 \quad ^4P^e_{3/2}$ 
& 1.627 & 1.634 & 2.446E13 & 2.566E13 \\ \cline{2-9}

& $K^{h2}_{\alpha}$ & 1.664$\pm$0.020
& $1s^02s^22p^4 \quad ^1S^e_{0}$ & $1s^12s^22p^3 \quad ^3S^o_{1}$ 
& 1.645 & 1.649 & 5.028E13 & 5.163E13 \\
&  &
& $1s^02s^22p^3 \quad ^4S^o_{3/2}$ & $1s^12s^22p^2 \quad ^4P^e_{3/2}$ 
& 1.661 & 1.662 & 2.054E13 & 2.037E13 \\ \cline{2-9} 

& $K^3_{\alpha}$ & 1.711$\pm$0.014
& $2p \quad ^2P^o_{3/2}$ & $1s \quad ^2S^e_{1/2}$ 
& 1.729 & 1.729 & 1.792E13 & 1.786E13 \\ \hline
{Ne} 
& $K_{\alpha \alpha}$ & 1.765$\pm$0.012
& $1s^02s^22p^6 \quad ^1S^o_{0}$ & $1s^22s2p^5 \quad ^2P^o_{1}$ 
& 1.765 & 1.758 & 5.930E9 & 7.045E9 \\ \hline

\end{tabular}}
\end{table*}

%%%%%%%%%%%%%%%%%%%%%%%%%%%%%%%%%%%%%%%%%%%%%%%%%%%%%%%%%%
\section{Results and discussion}
%%%%%%%%%%%%%%
\subsection{Spectral analysis}
Studying x-ray spectroscopy of light ions is very difficult, which can be understood from a comparison of the x-ray fluorescence yield as a function of the atomic number \cite{chen1980relativistic,chen1991auger}. We can see from the data tables given in these articles that the value of the Auger transition rate is much higher than the x-ray emission rate for $K$-shell transitions. The x-ray emission branches for Ne and Al are only 1.89\% and 4\%, respectively. Retardation of a plethora of Auger electrons in the target causes strong bremsstrahlung background to  the feeble x-ray signals. Furthermore, the large attenuation coefficient of low-energy x-rays through the detector window makes the x-ray detection too difficult. Thus, acquiring good statistics in the x-ray detections is extremely difficult with permissible accelerator timings. 
\par
X-ray spectra recorded for the Ne$^{6+}$ beam on the Al-target at 1.8 and 2.1 MeV are shown in Fig.\ref{fig:expt}. Fig.\ref{fig:expt}(A) shows the raw spectra at 1.8 MeV containing a prominent Bremsstrahlung background and some weakly spectral lines overriding this. Subtracting the background from the raw spectrum we obtain the net spectrum as shown in Fig.\ref{fig:expt}(B). The spectrum obtained shows a good peak structure belonging to Ne and Al in the region of 1.3 to 1.9 keV, which can be fitted into six Gaussian structures. Similar treatment was also applied on the raw spectrum obtained with the 2.1 MeV Ne$^{6+}$ beam. Here, we see a similar structure in the region of 1.3-1.9 keV along with a new structure appearing around 3.0 keV.
\par
The peak structure around 3.0 keV indicates its appearance as a result of the TEOP process in the Al-target atoms, which can be fitted into two Gaussian peaks. To know the correct energy of these peaks, we calibrated the spectra in a novel way. The bremsstrahlung backgrounds shown in both spectra have two absorption dips as marked in the figure. They appear due to the $K$-edge absorption of P \cite{franke1995p} and Cl \cite{kozimor2009trends} present as impurities in the Al-target foil. Since the $K$ absorption edges are very sharp and are known within the eV uncertainties, internal calibration done with these dips further refines the earlier calibration with the radioactive x-ray sources as mentioned above. With such a stringent calibration, the centroids of the Ne and Al x-ray lines are expected to be very correct. 
\par 
Though the spectra shown in Fig.\ref{fig:expt} do not show the Ne $K$ x rays due to their absorption in the mylar window, the TEOP lines originating in Ne projectile ions are observed to the right of the Al $K$ x ray peaks. To know the spectroscopic origin of the Ne as well as the Al TEOP lines we have made thorough calculations using the atomic structure codes GRASP and FAC as mentioned above. These calculations have been carried out according to the demand for the observed results in the present experiment, hence we have only considered the spectral lines that have measurable emission rates of the order of 10$^9$ s$^{-1}$ or more.
\par 
One important point is to mention here that in the present interactions, various ionic states can be possible not only with the projectile Ne ions but also with the Al-target atoms. Therefore, we have considered different charged states of both Ne and Al ions. Charge-state-dependent transition energies and rates have been calculated and listed in Table \ref{tab:theo}. The transition energies and rates have been compared with the available theoretical results. However, only the experimental transition energies are compared in Table \ref{tab:theo} as only the relative intensities are given for the experiments \cite{stoller1977two,auerhammer1988two} and are compared later. Comparison of the present calculation with the FAC code shows excellent agreement with the earlier theoretical calculations on both transition energies and rates except those given in \cite{mukherjee1990branching,mukherjee1997two}. This mismatch arises because the determination of the screening parameters was tuned here manually \cite{mukherjee1990branching,mukherjee1997two} rather than ab initio principles like other approaches, including the present methods. The results obtained with the GRASP code are also reasonably good. The scenario of comparison of our calculated transition energies with earlier experimental values is also very good. 
\par
\begin{table*}
\caption{\label{tab:Expt-2}Measured line energies with the 2.1 MeV Ne ions are listed and identified with the help of the present theoretical calculations as shown in Table \ref{tab:theo}. }  %
\resizebox{18cm}{!}{
\begin{tabular}{|l|l|l|l|l|l|l|l|l|}
\hline
El & Trans. & 2.1 MeV & Initial State & Final State & FAC & GRASP & A(FAC) & A(GRASP) \\ \hline
\multirow{5}{*}{Al} 
& $K^1_{\alpha}$ & 1.510$\pm$0.010 & $1s2s^22p^5 \ ^1P^o_1$ & $1s^22s^22p^4 \ ^3P^e_2$ & 1.502 & 1.503 & 2.738E13 & 2.833E13 \\
& & & $1s2s^22p^4\ ^4P^e_{5/2}$ & $1s^22s^22p^3\ ^2D^o_{5/2}$ & 1.514 & 1.513 & 1.463E13 & 1.284E13 \\ \cline{2-9}
& $K^2_{\alpha}$ & 1.572$\pm$0.012
& $1s2s^22p \quad ^1P^o_{1}$ & $1s^22s^2 \quad ^1S^e_{0}$ 
& 1.568 & 1.568 & 2.368E13 & 2.391E13 \\ 
& & & $1s2s2p \quad ^2P_{3/2}^{o}$ & $1s^22s \quad ^2S_{1/2}^{e}$ 
& 1.580 & 1.580 & 2.453E13 & 2.516E13 \\ \cline{2-9}
& $K^{h1}_{\alpha}$ & 1.611$\pm$0.026 & $1s^02s^22p^6 3s^2 3p \ ^2P_{1/2}^o$ & $1s^12s^22p^5 3s^2 3p\ ^2D_{3/2}^e$ & 1.613 & 1.619 & 2.995E13 & 3.552E13 \\
& & & $1s^02s^22p^6 3s^2 \ ^1S^e_0$ & $1s^12s^22p^5 3s^2\ ^1P^o_1$ & 1.614 & 1.621 & 5.187E13 & 6.188E13 \\
& & & $1s^02s^22p^6 3s \ ^2S^e_{1/2}$ & $1s^12s^22p^5 3s \ ^2P^o_{3/2}$ & 1.614 & 1.626 & 3.771E13 & 4.216E13 \\
& & & $1s^02s^22p^6 \ ^1S^e_0$ & $1s^12s^22p^5 \ ^1P^o_1$ & 1.615 & 1.627 & 6.144E13 & 6.548E13 \\ 
& & & $1s^02s^22p^5 \quad ^2P^o_{3/2}$ & $1s^12s^22p^4 \quad ^4P^e_{3/2}$ 
& 1.627 & 1.634 & 2.446E13 & 2.566E13 \\ \cline{2-9}
& $K^{h2}_{\alpha}$ & 1.664$\pm$0.026
& $1s^02s^22p^4 \quad ^1S^e_{0}$ & $1s^12s^22p^3 \quad ^3S^o_{1}$ 
& 1.645 & 1.649 & 5.028E13 & 5.163E13 \\
& & & $1s^02s^22p^3\ ^4S^0_{3/2}$ & $1s^12s^22p^2\ ^4P^e_{3/2}$ & 1.661 & 1.662 & 2.054E13 & 2.037E13 \\ \cline{2-9}
& $K^3_{\alpha}$ & 1.711$\pm$0.026
& $ 2p \quad ^2P^o_{3/2}$ & $1s \quad ^2S^e_{1/2}$ 
& 1.729 & 1.729 & 1.792E13 & 1.786E13 \\   \cline{2-9}
& $K^1_{\alpha \alpha}$ & 3.055$\pm$0.019 & $1s^02s^22p^63s^23p\ ^2P^o_{1/2}$ & $1s^22s2p^5 3s^2 3p\ ^2D^e_{3/2}$ & 3.058 & 3.072 & 5.139E10 & 7.620E10 \\
& & & $1s^02s^22p^6 3s^2 \ ^1S^e_{0}$ & $1s^22s2p^5 3s^2 \ ^3S^o_{0}$ & 3.059 & 3.057 & 1.066E11 & 1.707E11 \\
\cline{2-9}
& $K^2_{\alpha \alpha}$ & 3.117$\pm$ 0.022 & $1s^02s^22p^4\ ^1D^e_{2}$ & $1s^22s2p^3\ ^1P^0_{1}$ & 3.121 & 3.127 & 1.758E10 & 2.022E10 \\ \hline
{Ne} 
& $K_{\alpha \alpha}$ & 1.766 $\pm$ 0.020
& $1s^02s^22p^6 \quad ^1S^o_{0}$ & $1s^22s2p^5 \quad ^2P^o_{1}$ 
& 1.765 & 1.758 & 5.930E9 & 7.045E9 \\ \hline
\end{tabular}}
\end{table*}
%%%%%%%%%%%%%%%%%%%%%%%%%%%%%%%%%%%%%%%%%%%%%%
With the success mentioned above of our theoretical work, we decided to identify the observed spectral lines given in Table \ref{tab:Expt-1} for 1.8 MeV spectrum and Table \ref{tab:Expt-2} for 2.1 MeV spectrum with the help of our current calculations listed in Table \ref{tab:theo}. We can see that Table \ref{tab:Expt-1} lists only five lines of Al, including three $K_\alpha$ and two $K^h_\alpha$ lines, and one $K_{\alpha\alpha}$ line of Ne due to the collisions of the Ne ion on the Al-target at 1.8 MeV energy. On the other hand, we can see that Table \ref{tab:Expt-2} lists the six lines as observed with the 1.8 MeV experiment, in addition two TEOP lines of Al are also discerned with the 2.1 MeV experiment. In contrast, the Al ion on the Al-target at 25 MeV \cite{stoller1977two} reported a different set of 5 lines that included one $K_\alpha$ line, three $K^h_\alpha$ lines, and one $K_{\alpha\alpha}$ line. This difference can be attributed to different collision systems and also highly different impact energies. For the same reason, 20 keV electron beam impacting on an Al-target again shows a different set of fifteen emission lines including two $K_\alpha$, three $K^h_\alpha$, four $K_\beta$, and six $K_{\alpha\alpha}$ lines. Of these, only one line is common with our present work as a line $K_\alpha$ at 1.513 $\pm$ 0.003 keV, which we observe at 1.507 $\pm$ 0.011 keV and 1.510 $\pm$ 0.010 keV due to the impact of 1.8 and 2.1 MeV, respectively. Note though electron impact collisions evident $K_\beta$ lines, but neither \citet{stoller1977two} nor the present work exemplify such occurrences in the heavy-ion collisions.
%%%%%%%%%%%%%%%%%%%%%%%%%%%%
\subsection{Relative intensity}
Transition probabilities have been studied theoretically along with the transition energies so that the predictive power of the theoretical calculations can be tested better if they are compared well with the measured values. To do so, we first compare the ratio of the transition rates with the previously measured experimental intensity ratio of $I(K_{\alpha\alpha})/I(K_\alpha^h)$. Nevertheless, we failed to use the experimental result of \citet{stoller1977two} as it lists three $K^h_\alpha$ lines and one $K_{\alpha\alpha}$ line, but reports only one relative intensity. Here, which $K^h_\alpha$ line was taken into account in determining the relative intensity is not mentioned in the discussion.  Such problem is not encountered in \citet{auerhammer1988two}. The measured ratio $I(K_{\alpha\alpha})/I(K_\alpha^h)$=(2.2$\pm$0.8)$\times$$10^{-3}$ \cite{auerhammer1988two} agrees very well with the predictions of FAC ($1.72\times 10^{-3}$) and GRASP (2.15$\times 10^{-3}$). Here, $K_{\alpha\alpha}$ and $K_\alpha^h$ lines represent $1s^02s^22p^6 3s^23p \quad     ^2P^{o}_{1/2}$ - $1s^22s2p^5 3s^2 3p \quad ^2D^{e}_{3/2}$ and $1s^02s^22p^6 3s^23p \quad ^2P_{1/2}^{o}$ - $1s^12s^22p^5 3s^2 3p\quad ^2D_{3/2}^{e}$ transitions, respectively. This ensures the good predictive power of the theories used here.
\par
Let us next take the present case.  Here, the $K_\alpha$ and $K^h_\alpha$ lines appear around the Al $K$-edge region (1.566 keV), and thus it is difficult to find the quantum efficiency of the detector correctly. In contrast, the Al $K_{\alpha\alpha}$ lines fall in a safe region, and thus the measurement uncertainty due to the quantum efficiency is not at all doubtful. The SDD efficiency for the detector used was measured and reported in \citet{ahmad2022inner}, we can get the efficiency from Fig. 4 of it. The measured intensity ratio of $K^1_{\alpha\alpha}$ to $K^2_{\alpha\alpha}$  for Al is 3.04$\pm$ 0.28 as given in Table \ref{tab:IR}. This measured ratio gives us a correct clue as to which transition among the two possibilities given in Table \ref{tab:Expt-2} can be attributed to $K^1_{\alpha\alpha}$. If we assign the first one ($1s^02s^22p^63s^23p\ ^2P^o_{1/2}$ - $1s^22s2p^5 3s^2 3p\ ^2D^e_{3/2}$) to it, the theoretical ratios predicted by both the theories agree well (Fac and GRASP predicts 2.92 and 3.7, respectively) with the measurement. Hence, consideration of transition energies and the transition rates helps us identify the origin of the observed spectral lines quite well. 
\par
If a certain $K_{\alpha\alpha}$ line is observed, its complimentary $K^h_{\alpha}$ line must also be seen in the spectrum. In other words, a pair of observed $K^{h1}_{\alpha}$ and $K^1_{\alpha\alpha}$ lines must have the same lower level. Since the $1s^02s^22p^63s^23p\ ^2P^o_{1/2}$ level is the lower level of $K^1_{\alpha\alpha}$, it must also be the lower level of the $K^{h1}_{\alpha}$ line. Now it is noteworthy that, though five transitions (see Table \ref{tab:Expt-2}) are possible to constitute the $K^{h1}_{\alpha}$ line as far as the closeness of the experimental transition energy with the theoretically obtained ones is concerned. Nevertheless, only one transition contains the lower level mentioned above $1s^02s^22p^63s^23p\ ^2P^o_{1/2}$. Thus, the $K^{h1}_{\alpha}$ line is due to the transition $1s^02s^22p^63s^23p\ ^2P^o_{1/2}$ - $1s^12s^22p^5 3s^2 3p\ ^2D_{3/2}^e$. Similar arguments fix the other pair of $K^2_{\alpha\alpha}$ and $K^{h2}_\alpha$ lines as given in Table \ref{tab:Expt-3}.  However, it is not possible to fix the final transition for the $K_\alpha$ lines as they do not have any connection either with the $K^{h}_{\alpha}$ or $K_{\alpha\alpha}$ line.
\par
Next, to compare the measured intensity ratios between the OEOP and TEOP lines emerging from the target, we are bound to use the SDD efficiencies measured in our laboratory. This time we use the fitted efficiency equation given in Ref. \cite{ahmad2022inner}.  The ratio between the OEOP and TEOP line intensities gives us a striking surprise: the measured intensity ratio $\frac{I(K^1_\alpha+K^2_\alpha+K^3_\alpha)}{I(K^1_{\alpha\alpha}+K^2_{\alpha\alpha})}=82$ is about three times smaller than the theoretical intensity ratio ($\frac{I(1s2s^22p^6 3s^23p \quad^3P_0^o-1s^22s^22p^5 3s^23p  \quad ^3D_1^e)}{I(1s^02s^22p^63s^23p\ ^2P^o_{1/2}-1s^22s2p^5 3s^2 3p\ ^2D^e_{3/2})}=2.47\times10^2$. It implies that either the population of $K_\alpha$ is highly reduced or that of $K_{\alpha\alpha}$ is greatly enhanced, or both the phenomena are taking place simultaneously. To the best of our knowledge, no such physical occurrence is familiar to us till date. Nevertheless, the observed fact must have a certain truth and we attempt to explore it.
\par
%%%%%%%%%%
%%%%%%%%%%%%%%
\begin{table}
\caption{\label{tab:IR} Comparison of the measured line intensity ratios with the calculated ones  in Ne and Al ions.}
\begin{tabular}{l l l l l l l l l l l l l l l l}
\hline
 \makecell{El} &  \makecell{Ratio} & \makecell{2.1 MeV}&   \makecell{FAC}&   \makecell{GRASP} \\
Al& $\frac{{{K^1_{\alpha\alpha}}}}{{{K^2_{\alpha\alpha}}}}$  & \makecell{3.04$\pm$0.28}& \makecell{2.92}& \makecell{3.7}\\
\hline
\end{tabular}
\end{table}
%%%
%%
\begin{table*}
\caption{\label{tab:Expt-3} Measured line energies with the 2.1 MeV Ne ions are listed and identified with the help of the theoretical calculations presented here as shown in Table \ref{tab:theo}. }  %
\resizebox{17cm}{!}{
\begin{tabular}{|l|l|l|l|l|l|l|l|l|}
\hline
El & Trans. & 2.1 MeV & Initial State & Final State & FAC & GRASP & A(FAC) & A(GRASP) \\ \hline
\multirow{5}{*}{Al} 
& $K^{h1}_{\alpha}$ & 1.611$\pm$0.026 & $1s^02s^22p^6 3s^2 3p \ ^2P_{1/2}^o$ & $1s^12s^22p^5 3s^2 3p\ ^2D_{3/2}^e$ & 1.613 & 1.619 & 2.995E13 & 3.552E13 \\
 \cline{2-9}
& $K^{h2}_{\alpha}$ & 1.664$\pm$0.026
& $1s^02s^22p^4 \quad ^1S^e_{0}$ & $1s^12s^22p^3 \quad ^3S^o_{1}$ 
& 1.645 & 1.649 & 5.028E13 & 5.163E13 \\
 \cline{2-9}
& $K^1_{\alpha \alpha}$ & 3.055$\pm$0.019 & $1s^02s^22p^63s^23p\ ^2P^o_{1/2}$ & $1s^22s2p^5 3s^2 3p\ ^2D^e_{3/2}$ & 3.058 & 3.072 & 5.139E10 & 7.620E10 \\
\cline{2-9}
& $K^2_{\alpha \alpha}$ & 3.118$\pm$ 0.022 & $1s^02s^22p^4\ ^1D^e_{2}$ & $1s^22s2p^3\ ^1P^0_{1}$ & 3.121 & 3.127 & 1.758E10 & 2.022E10 \\ \hline
\end{tabular}}
\end{table*}
%%%%%%%%%%%%%%%%%%%%%%
\subsection{Effect of bremsstrahlung radiation in Double $K$-shell vacancy production}
As mentioned above, a remarkable feature is observed in the present experiment: The population of $K_{\alpha}$ and $K_{\alpha\alpha}$ is highly altered at low-energy collisions due a hitherto unknown process. First of all, observation of $K_{\alpha\alpha}$ line at such a low-energy collision is highly unlikely due to collisional shake down model \cite{aaberg1976origin}. Truth of this statement can be realized from the scenario reported in \citet{stoller1977two}. We can find there that a mild signature of $K_{\alpha\alpha}$ lines exists with 40 MeV Fe beam impacting on Ni-target, but such occurrence is fully disappeared while beam energy is reduced to 25 MeV. Another example may be quoted for a collision system of Cl-beam on Ni-target. For this case the $K_{\alpha\alpha}$ lines are unobserved with both 25 and 40 MeV energies. These facts suggest that no way 2.1 MeV beam can cause appearance of $K_{\alpha\alpha}$ lines on account of the collisional shake down mechanism. It raises thus a big challenge to unearth the mechanism behind the copious population of the $K_{\alpha\alpha}$ lines at such a low energy (105 keV/u Ne beam on Al-target). 
\par
We have observed the prominent bremsstrahlung backgrounds and two absorption dips in Fig. \ref{fig:expt}. The bremsstrahlung radiation causes these absorption dips appearing due to the $K$-edge absorption of P and Cl present as impurities in the Al-target foils. The bremsstrahlung radiation induced absorption dips take place with impurity atoms, the bremsstrahlung photon can also be absorbed by the Al-target atoms as well as projectile ions, which may lead to photo excitation as well as photoionization in both target atoms and projectile ions. Such an event is found to occur when an electron beam is passed through  a plasma and the bremsstrahlung radiation created in the collisions causes runaway electrons to gain energy \cite{bakhtiari2005role}. It is extremely difficult to distinguish the bremsstrahlung photon induced ionization from those produced by the ion-solid collision itself. In search of such an event, we encounter the signature of the $K_{\alpha\alpha}$ lines, which are not the outcome of the present collision processes. It will be highly intriguing to explore the effect of bremsstrahlung radiation in producing the $K_{\alpha\alpha}$ lines.
\par
Past works, for example \citet{khan1965studies}, evidence that the projectiles with keV energies can create $K$ x-rays and we observe such emissions in this present experiment too. We have applied theoretical approach similar to  \citet{kaur2023understanding} to estimate the x-ray production cross sections for the collision of 105 keV/u Ne beam on Al-target. This theory \cite{kaur2025toward} takes account of the major processes significant for heavy ion collision induced x-ray emissions such as multiple ionization effects in the outer shells and electron capture of target inner-shell electrons by the projectile ions so correctly that theoretical estimates are well aligned with experimentally measured x-ray production cross sections \cite{chatterjee2021significance,kaur2025toward}. 
\par
The estimated $K$ x-ray production cross section for the collision of 105 keV/u Ne beam on Al-target turns out to be 1999 mb. In this calculation correct values of $K$ x-ray fluorescence yield and Fermi velocity are important and the used figures for these are 0.0365$\pm$0.0025 \cite{kaur2025toward} and 1.599 $\times 10^6$ m/sec \cite{gall2016electron}, respectively. In order to compare this estimated $K$ x-ray production cross sections with the measured ones,
we sum up all the three $K_\alpha$ (See Table III) line intensities together as the total $K_\alpha$ line intensity. With this consideration, the $K$ x-ray production cross section turns out only 360$\pm$42 mb. This theoretical estimate is very accurate, and we expect close agreement with the experiment. But the measured cross section is only 0.18 $\times$ the theoretical cross section.  
\par
\begin{figure}
    \centering
    \includegraphics[width=1.00\linewidth]{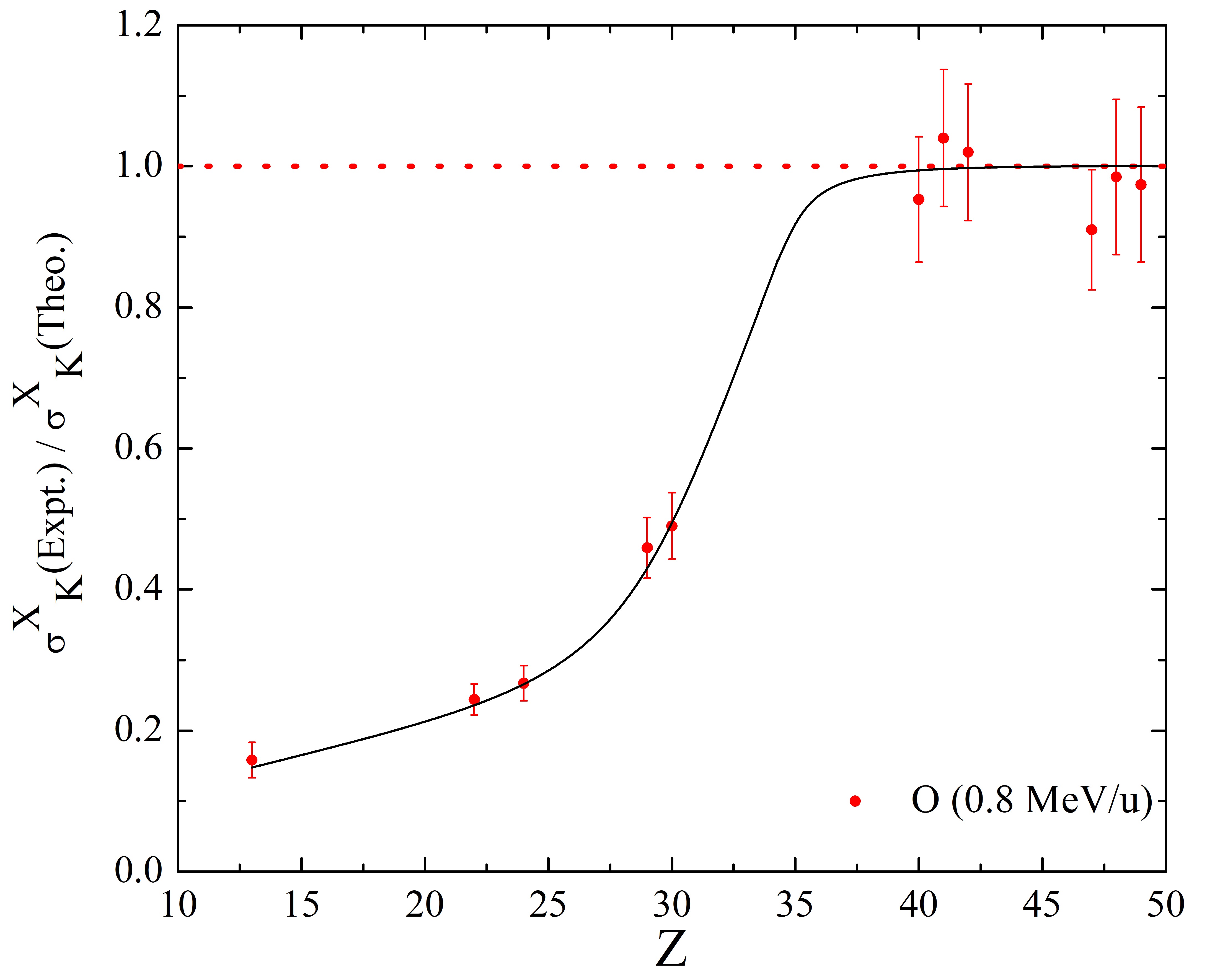}
    \caption{Experimental to theoretical $K$ x-ray production cross section ratio is plotted for 0.8 MeV/u oxygen ion impacting on various targets designated by their atomic number Z. The experimental data are taken from \citet{gorlachev2016k}. The solid line is drawn to indicate a possible trend.}
    \label{fig:goralchev}
\end{figure}

The above observation seems to indicate a new physical process in an area of physics research, which has remained active since the 1970s \cite{brandt1973dynamic}. Hence, the said type of scenarios must also be found in the earlier heavy-ion collision-induced $K$ X-ray production cross-section measurements. With such expectations, we came across many experiments \cite{gorlachev2016k,gluchshenko2016k,gorlachev2017k,gorlachev2018k} indicating similar truths. The measured $K$ x-ray production cross sections are seen to be much lower than the theoretical calculations. We investigated a case where an oxygen ion collided with various targets, including an Al-target \cite{gorlachev2016k}. We have plotted the ratio of measured K x-ray production cross section ($\sigma^x_K$) by the theoretically calculated $\sigma^x_K$ as a function of the atomic number of the target elements and shown in Fig. \ref{fig:goralchev}. Here, we can see how large this disagreement can be for the light target elements. Interestingly, this mismatch keeps on reducing as we move from lighter (Al-target) to the heavier (Z-target) targets. %Such a scenario is shown in Fig. \ref{fig:goralchev}. 
\par
Above mentioned reduction of the $K$ x-ray production cross section in earlier experiments may imply to the authors that the theoretical calculations may overestimate the heavy-ion collision-induced $K$ x-ray production cross sections because they studied only the $K$ x-ray emissions. In contrast, we have investigated the $K_{\alpha\alpha}$ x-rays along with the $K_\alpha$ x-rays, where we notice a reduction of the $K$ x-ray yields and enhancement of the $K_{\alpha\alpha}$ x-ray production. Furthermore, we mentioned above that the $K_{\alpha\alpha}$ x-ray production is not at all eminent due to atomic collisions at such a low energy. This unusual observation suggests that a large portion of the single $K$ shell vacant states is altered prior to the $K$ x-ray emission to some other type of state, especially the doubly $K$-shell vacant states so that the $K_{\alpha\alpha}$ x-rays are evidenced in the experiment. Hence, the ratio of the measured $\sigma^x_K$ by the theoretically calculated $\sigma^x_K$ gives a measure of the survival probability of the $K$ x-ray yields. Let us define this ratio as the survival probability of the $K_\alpha$ x-rays (P$_{\text{sur}}$) as follows:
\begin{equation}
    P_{\text{sur}}=\frac{\text{estimated}~\sigma^x_K}{\text{measured}~\sigma^x_K}
    \label{Eq:P_sur}
\end{equation}
\noindent This means that Fig. \ref{fig:goralchev} shows a P$_{\text{sur}}$ vs target atomic number for the impact of oxygen ions of 0.8 MeV/u. P$_{\text{sur}}$ increases to unity with an increasing target atomic number. In other words, the probability of $K_\alpha$ x-ray annihilation (1-P$_{\text{sur}}$) decreases to zero with increasing target atomic number. Therefore, Eqn.\ref{Eq:P_sur} gives a measure of P$_{\text{sur}}$ for neon-aluminum collisions at 2.1 MeV; it is only 0.18. Our curiosity is to know now what physical origin is responsible for such a phenomenon. 
\par
A few interesting characteristics of the bremsstrahlung background and its effect on the observed spectra have been studied in Table \ref{tab:Brems}. We can notice here that the normalized (with respect to the same charge counts collected in the current integrator at different beam energies) bremsstrahlung x-ray intensity ratio at two beam energies 1.8 and 2.1 Mev is 0.745$\pm$0.010 and it is directly reflected to the absorption $K$-edge intensity ratio of a particular impurity atom, for P (0.752$\pm$0.008) and for Cl (0.751$\pm$0.007). Furthermore, the absorption $K$ edge intensity ratio of the two impurity atoms remains constant at every beam energy as the ratio is 1.195$\pm$0.010 and 1.202$\pm$0.011 for 1.8 and 2.1 MeV, respectively. However, visual inspection on the x-ray yields at two beam energies 1.8 and 2.1 Mev in comparison to the corresponding bremsstrahlung backgrounds gives a different picture; relative yield of x-rays with respect to the bremsstrahlung radiation at 1.8 Mev looks smaller than that at 2.1 MeV. We found in fact that it is true since the total x-ray count in the region of 1.3-1.9 keV by the total bremsstrahlung background ratio at 1.8 MeV is 0.929$\pm$0.010, while the same at 2.1 MeV is 1.209$\pm$0.011. %\textcolor{red}{If we compare the same x-ray counts with respect to the K absorption edge of P, we get similar results as shown in Table \ref{tab:Brems}.} 
\par 
Let us assume that the bremsstrahlung radiation alters a large portion of the $K_\alpha$ x-ray yield to produce the whole of the observed $K^h_\alpha$ and $K_{\alpha\alpha}$ x-ray emissions in the experiment. It means the $K_\alpha$ x ray is the sole origin of all the observed $K^h_\alpha$ and $K_{\alpha\alpha}$ x-ray lines. In other words, Al-$K_\alpha$ cross section only governs the total x-ray yield (excluding the sixth or Ne $K_{\alpha\alpha}$ peak given in Table II and III, which is related to Ne $K_{\alpha}$) of Al-$K^h_\alpha$ and Al-$K_{\alpha\alpha}$ cross sections. Therefore, the weaker x-ray yield at 1.8 MeV than that at 2.1 Mev is related to the following factors only: (i) the $K$ x-ray production cross section ($\sigma^x_K$) at 1.8 MeV (815 mb) is smaller than that at 2.1 MeV energy (1999 mb) and (ii) difference in P$_{\text{sur}}$ of the $K$ x-rays (Eqn.\ref{Eq:P_sur}) in the two beam energies. P$_{\text{sur}}$ for the 2.1 MeV case is about 0.18. The same at 1.8 Mev is found to be about 0.14 as the measured cross section is only 115 mb with respect to the correct theoretical cross section 815 mb. These two factors combined give rise to the x-ray yield ratio R$^x_Y$ as follows:
\begin{equation}
R^x_Y=\frac{\text{$\sigma^x_K$ at 1.8 MeV}\times\text{$P_{\text{sur}}$ at 1.8 MeV}}{\text{$\sigma^x_K$ at 2.1 MeV}\times\text{$P_{\text{sur}}$ at 2.1 MeV}}=0.317.
\label{Eq:Rxy}
\end{equation}
\noindent This estimated $R^x_Y$ is extremely close to the measured value o.315$\pm$0.008 as shown in Table \ref{tab:Brems}. Thus, we proved here that the above assumption is in fact a physical phenomenon by which most $K_\alpha$ x-rays are destroyed due to bremsstrahlung radiation-induced photoionization at low energies.
\par 
%%%%shifted at the end%%%We mentioned above that the present experimental condition has potential to produce single $K$-shell vacancies in the target atoms. If the states with single $K$-shell vacancy (lower level of $K$ x-ray lines) can absorb the bremsstrahlung x-ray photons of appropriate energy, so that the states with double $K$-shell vacancy are produced and subsequently the $K_{\alpha\alpha}$ emissions can emerge. However, such an occurrence is seen at only 2.1 MeV experiments. We have observed certain characteristics that support this observation very well. The normalized bremsstrahlung yield at 1.8 MeV is 0.745 times that at 2.1 MeV (Table \ref {tab:Brems}). Furthermore, theoretically estimated $K$ x-ray production cross section at 1.8 MeV is 0.815 b and that at 2.1 MeV is 1.999 b. Both of these factors suggest that the 2.1 MeV Ne has a higher probability of producing the $K_{\alpha\alpha}$ x-ray emission than the 1.8 MeV energy. Furthermore, the population of $K_{\alpha\alpha}$ x-ray line has to be good enough to show it up in the spectrum, as the transition rate of the $K_{\alpha\alpha}$ line is at least two orders of magnitude (2.47$\times 10^2$) lower than that of the $K_{\alpha}$ line, as mentioned above. All these tree factors led us not to observe the Al $K_{\alpha\alpha}$ line in 1.8 MeV experiment.
\par
The energy of the above mentioned bremsstrahlung x-ray photons that causes the $K_{\alpha\alpha}$ x-ray emission in the low energy ion-solid collisions is equal to the energy difference between the $K_{\alpha\alpha}$  and $K_{\alpha}$ x-rays. In the present case, the energy difference between the $K^1_{\alpha\alpha}$  and $K^1_{\alpha}$ x-rays and that between the $K^2_{\alpha\alpha}$  and $K^2_{\alpha}$ x-rays is incidentally the same 1.545 keV. This means that the $K^1_{\alpha}$ and $K^2_{\alpha}$ x-rays are promoted to $K^1_{\alpha\alpha}$  and $K^2_{\alpha\alpha}$, respectively. In such cases, the bremsstrahlung x-ray photon energies $\geq$ 1.545 keV can ionize aluminum atoms having single $K$-shell vacancies to aluminum atoms with double $K$-shell vacancies. Since any bremsstrahlung x-ray photon having energies $\geq$ 1.545 keV can participate in the said process, therefore no experimental signature can be found in the bremsstrahlung profile. 
\par
\begin{table}
\caption{Various intensity ratios to understand the Bremsstrahlung photon induced photoionization phenomena.}  %
\resizebox{8.8cm}{!}{
\begin{tabular}{|cl|l|}
\hline
\multicolumn{2}{|c|}{Intensity ratio for}                                                                                                         & Ratio \\ \hline
\multicolumn{2}{|c|}{\begin{tabular}[c]{@{}c@{}}Normalized bremsstrahlung peaks at 1.8 and 2.1 MeV\end{tabular}} & 0.745$\pm$0.010  \\ \hline
\multicolumn{2}{|c|}{Absorption peak intensity of P at 1.8 and 2.1 MeV}                                    & 0.752$\pm$0.008  \\ \hline
\multicolumn{2}{|c|}{Absorption peak intensity of Cl at 1.8 and 2.1 MeV}                                    & 0.751$\pm$0.007  \\ \hline
\multicolumn{2}{|c|}{\begin{tabular}[c]{@{}c@{}}Absorption peak intensity of P and Cl at 1.8 MeV\end{tabular}} & 1.195$\pm$0.010  \\ \hline
\multicolumn{2}{|c|}{\begin{tabular}[c]{@{}c@{}}Absorption peak intensity of P and Cl at 2.1 MeV\end{tabular}} & 1.202$\pm$0.011  \\ \hline
\multicolumn{2}{|c|}{\begin{tabular}[c]{@{}c@{}}K x-ray peak and bremsstrahlung peak at 1.8 MeV\end{tabular}} & 0.929$\pm$0.010 \\ \hline
\multicolumn{2}{|c|}{\begin{tabular}[c]{@{}c@{}}K x-ray peak and bremsstrahlung peak at 2.1 MeV\end{tabular}} & 1.209$\pm$0.011\\ \hline
%\multicolumn{2}{|c|}{\begin{tabular}[c]{@{}c@{}}K x-ray peak and absorption peak of P at 1.8 MeV\end{tabular}} & 16.726$\pm$0.181\\ \hline
%\multicolumn{2}{|c|}{\begin{tabular}[c]{@{}c@{}}K x-ray peak and absorption peak of P at 2.1 MeV\end{tabular}} & 21.038$\pm$0.212\\ \hline
\multicolumn{2}{|c|}{Normalized K x-ray peaks at 1.8 and 2.1 MeV}                                  & 0.315$\pm$0.008  \\ \hline
\end{tabular}}
\label{tab:Brems}
\end{table}
%%%%%%%%%%%%%%%%%%%%%%%%%%%%%%%%%
\par
%%%%%%%%%%%%%%%%%%5
\textcolor{black}{In another attempt, we intend to check whether the whole of annihilated singly vacant $K$ shell states can fully be converted to doubly vacant K shell states leading to K$^h_\alpha$ and K$_{\alpha\alpha}$ x-ray emissions. Such a chance is highly unlikely; rather, it can have a certain conversion probability. We call it the single to double $K$ vacancy conversion probability ($P_{\text{con}}$) and it is written as 
\begin{equation}
P_{\text{con}}=\frac{(K^h_\alpha+K_{\alpha\alpha})~ \text{yield}}{\text{Annihilated}~  K_\alpha~ \text{yield}}=0.0966%\times{\text{relative rate}}=0.66.
\label{Eq:P_con}
\end{equation}
\noindent Here, $\text{annihilated}~ K_\alpha~ \text{yield}=(1-P_{\text{sur}})\times \text{total}~ K_\alpha~ \text{yield}$. Here, $\text{total}~ K_\alpha~ \text{yield}=\frac{\text{measured}~ K_\alpha~ \text{yield}}{0.18}$. Thus, only 9.66\% of the annihilated $K_\alpha$ yield is converted to the K$^h_\alpha$ and K$_{\alpha\alpha}$ lines.} Note that the annihilated $K_\alpha$ yield is predominantly converted mainly to the K$^h_\alpha$ yield, since the measured branching ratio of the K$_{\alpha\alpha}$ line ($\frac{\text{Intensity of~} K_{\alpha\alpha}}{\text{total intensity of~} K^h_\alpha \text{~and~} K_{\alpha\alpha}}$) in the 2.1 MeV experiment is only 0.027. Since the branching ratio is very small, even though the K$_{\alpha\alpha}$ lines were not observed in the 1.8 MeV experiment, we can evaluate $P_{\text{con}}$ approximately ignoring the small contribution of the K$_{\alpha\alpha}$ line intensity in the estimation of $P_{\text{con}}$ using Eqn.\ref{Eq:P_con}. We find it to be only 0.0455. This gives a concrete reason for not observing the K$_{\alpha\alpha}$ lines in the 1.8 MeV experiment.
\par
We mentioned above that the present experimental condition has potential to produce single $K$-shell vacancies in the target atoms. If the states with single $K$-shell vacancy (lower level of $K$ x-ray lines) can absorb the bremsstrahlung x-ray photons of appropriate energy, so that the states with double $K$-shell vacancy are produced and subsequently the $K_{\alpha\alpha}$ emissions can emerge. However, such an occurrence is seen at only 2.1 MeV experiments. We have observed certain characteristics that support this observation very well. (i) The normalized bremsstrahlung yield at 1.8 MeV is 0.745 times that at 2.1 MeV (Table \ref {tab:Brems}). (ii) Theoretical $K$ x-ray production cross section at 1.8 MeV is 0.815 b and that at 2.1 MeV is 1.999 b. (iii) The single to double $K$ vacancy conversion probability $P_{\text{con}}$ for 1.8 and 2.1 MeV is 0.0455 and 0.0966, respectively. All these three factors suggest that the probability of producing the $K_{\alpha\alpha}$ x-ray emission at 2.1 MeV is 6.99 times higher than the 1.8 MeV energy. Furthermore, the population of $K_{\alpha\alpha}$ x-ray line has to be good enough to show it up in the spectrum, as the transition rate of the $K_{\alpha\alpha}$ line is at least two orders of magnitude (2.47$\times 10^2$) lower than that of the $K_{\alpha}$ line, as mentioned above. This factor proves that the Al $K^h_{\alpha}$ lines are observed, but the Al $K_{\alpha\alpha}$ lines are not so in 1.8 MeV experiment.
\par
In a final attempt, we wanted to validate the phenomenon of generating double K shell vacancies induced by bremsstrahlung from single K shell vacancies produced in solid-ion collisions. Interestingly, we see that the ratio of the intensity sum of three $K_{\alpha}$ peaks at 1.8 MeV to that of 2.1 MeV is exactly equal (within 0.2\%) to the ratio of the intensity sum of the five peaks including three K$_{\alpha}$ and two K$^h_{\alpha}$ peaks at 1.8 MeV (Table II) by that of 2.1 MeV (Table III). This means that the intensity of the two K$^h_{\alpha}$ peaks is directly proportional to the sum of the intensity of the three K$_{\alpha}$ peaks at every beam energy (1.8 or 2.1 MeV). The proportionality constant in the 1.8 MeV case is equal (within 0. 5\%) to that of the 2.1 MeV case. Physically, this implies that the K$^h_{\alpha}$ and K$_{\alpha\alpha}$ lines originate in fact from the K$_{\alpha}$ lines. More correctly, the singly ionized K-shells are converted to doubly ionized K-shells, so that instead of K$_{\alpha}$ x-rays, the K$^h_{\alpha}$ and K$_{\alpha\alpha}$ x-rays emerge. Although it confirms that singly-ionized K-shells are converted to doubly ionized K-shells, however, it does not ensure the conversion mechanism. The only viable mechanism responsible is the bremsstrahlung radiation induced photoionization discussed above.%The proportionality constant is related to a product of the survival probability and the singly vacant K shell states to doubly vacant K shell states. }
%%%%%%%%%%%%%%%%%%%%%%
%%%%%%%%%%%%%%%%%%%%%%%%%%%%%%%%%%%%%%%%%%%%%%%%%
%%%%%%%%%%%%%%%%%%%%%%%%%%%%%%%%%%%%%%%%%%%%%%%%%
\section{Conclusion}
To conclude, we have succeeded in segregating the TEOP spectra from prominent Bremsstrahlung backgrounds in the interaction of slow neon ions on an Al-target (90-105 keV/amu). The correct energy of the TEOP lines is measured by incorporating the internal energy calibration using the $K$-absorption edges of the impurities of P and Cl present in the Al-target. To assign such TEOP lines, we have performed extensive calculations using the atomic structure codes GRASP and FAC. The calculations performed show excellent agreement with the earlier experimental results. This assignment has also been verified with the comparison between the measured and calculated transition rates. Most importantly, the excitation mechanism at the low energies suggest that although the present experimental condition (2.1 MeV Ne on Al-target) can produce $K_{\alpha}$ x-ray lines, but cannot create $K_{\alpha\alpha}$ x-ray lines at all, the $K_{\alpha\alpha}$ x-ray lines are in fact built due to two favorable reasons: (i) the majority ($\approx 82\%$) of the $K_\alpha$ yield is annihilated due to bremsstrahlung radiation induced photoionization and (ii) This annihilated $K_\alpha$ yield converts to the $K_{\alpha\alpha}$ emissions with a large probability 0.0966. The K x-ray production cross section and K x-ray survival probability describe very well the total $K$ x-ray yield measured at the two beam energies 1.8 and 2.1 MeV. Another interesting observation is that the 1.8 MeV Ne ion on the Al-target does not show any TEOP lines of Al, whereas the 2.1 MeV Ne ion does. The reason behind it has been well described in terms of lower $K$ x-ray production cross sections, lower bremsstrahlung yields and lower single to double $K$ vacancy conversion probability at 1.8 MeV than those at 2.1 MeV. Thus, above reasons reveal well the proposed bremsstrahlung radiation induced photoionization mechanism. 
\par
This new Bremsstrahlung photon induced photoionization process may revolutionize the various fields of research such as plasma physics, astronomy, and astrophysics. For example, an anomalously bright line is observed in Perseus at 3.62 keV \cite{bulbul2014detection}. It is close to an Ar XVII dielectronic recombination line; however, its emissivity is not at all matched with the expected value and physically difficult to understand. The observed spectra contain prominent silicon $K_\alpha$ line and strong bremsstrahlung radiation. Furthermore, the line energy is quite close to a TEOP line of Silicon estimated at 3.589 keV \cite{kadrekar2010two}. Hence, the 3.62 keV line might belong to a silicon TEOP line.

%%%%%%%%%%%%%%%%%%%%%%%%%%%%%%%%%%%%%
\begin{acknowledgments}
The authors are extremely grateful to Prashant Sharma for his invaluable comments.
The authors thank the LEIBF facility crew of IUAC consisting of Krishna Kant Pal, Amit Kumar, and Pravin Kumar. They are also grateful to S.R. Abhilash and D. Kabiraj for providing the necessary facilities and support to make thin film targets. In addition, Narendra Kumar expresses his gratitude to the Department of Science and Technology (DST) for awarding him Inspire-Senior Research Fellowship, award no. IF210224. Alok Kumar Singh Jha gratefully acknowledges the financial assistance provided by SERB, Department of Science and Technology (DST), Government of India under research grant no. CRG/2022/008061.
\end{acknowledgments}
\nocite{*}
\bibliography{apssamp}% Produces the bibliography via BibTeX.
\end{document}